\documentclass[11pt]{elsarticle}
\usepackage{latexsym}
\usepackage{amssymb}
\usepackage{amsmath}
\usepackage{amsthm}
\usepackage{amsfonts}
\usepackage{amstext}
\usepackage{multicol,multirow}
\usepackage{rotating}
\usepackage{graphicx}
\usepackage{caption}
\usepackage{subcaption}
\usepackage{url}
\usepackage{lineno}
\usepackage[margin=1in]{geometry}
\usepackage{charter}
\usepackage{parskip}
\newcounter{ordprop}

\newtheorem{proposition}[ordprop]{Proposition}

\begin{document}
\journal{Water Research}

\begin{frontmatter}

\title{Persistence--infectivity trade-offs in environmentally transmitted pathogens change population-level disease dynamics}
\author[1]{Andrew F. Brouwer}\corref{mycorrespondingauthor}
\cortext[mycorrespondingauthor]{Corresponding author}
\ead{brouweaf@umich.edu}
\author[1]{Marisa C. Eisenberg}
\author[1]{Nancy G. Love}
\author[1]{Joseph N. S. Eisenberg}
\address[1]{Department of Epidemiology, University of Michigan, 1415 Washington Heights, Ann Arbor, MI 48109}
\address[2]{Department of Civil and Environmental Engineering, University of Michigan, 1351 Beal Avenue, Ann Arbor, MI 48109}

\date{}

\begin{abstract}
\noindent
Human pathogens transmitted through environmental pathways are subject to stress and pressures outside of the host. These pressures may cause pathogen pathovars to diverge in their environmental persistence and their infectivity on an evolutionary time-scale. On a shorter time-scale, a single-genotype pathogen population may display wide variation in persistence times and exhibit biphasic decay. Using an infectious disease transmission modeling framework, we demonstrate in both cases that fitness-preserving trade-offs have implications for the dynamics of associated epidemics: less infectious, more persistent pathogens cause epidemics to progress more slowly than more infectious, less persistent (labile) pathogens, even when the overall risk is the same. Using identifiability analysis, we show that the usual disease surveillance data does not sufficiently inform these underlying pathogen population dynamics, even with basic environmental monitoring. These results suggest directions for future microbial research and environmental monitoring. In particular, determining the relative infectivity of persistent pathogen subpopulations and the rates of phenotypic conversion will help ascertain how much disease risk is associated with the long tails of biphasic decay. Alternatively, risk can be indirectly ascertained by developing methods to separately monitor labile and persistent subpopulations. A better understanding of persistence--infectivity trade-offs and associated dynamics can improve risk assessment and disease control strategies.
\end{abstract}


\end{frontmatter}

%\begin{figure}[htbp]
%\centering
%
%    \includegraphics[width=1\textwidth]{Figures/Graphical_abstract}
%\captionsetup{labelformat=empty}
%    \caption{Graphical abstract}
%    \label{}
%\end{figure}
%\setcounter{figure}{0}  
%
%\section*{Highlights}
%\begin{itemize}
%\item Pathogens make trade-offs in response to environmental stress 
%\item We model a trade-off between environmental persistence and infectivity 
%\item The trade-off can preserve the attack ratio but result in different epidemic dynamics 
%\item More persistent, less infectious pathogens cause slower epidemics
%\item Targeted experiments could maximize information from disease surveillance data 
%\end{itemize}

\clearpage
%\linenumbers
\section{Introduction}
Many human pathogens, particularly waterborne enteric pathogens, require a host to reproduce but are transmitted through the environment where they are subjected to a variety of stressors. These stressors can result in long-term selection pressure that causes pathogens to evolve over time; e.g., the transmission--virulence trade-off~\citep{Alizon2009} and virulence--environmental persistence trade-off~\citep{Walther2004} are thought to arise from evolutionary pressures. Short-term environmental stress, on the other hand, can lead to different dynamics in different environmental conditions; e.g., changes in kinetics~\citep{Pepper2015}, morphology~\citep{Adams2003}, or antimicrobial resistance~\citep{Poole2012} can occur on this time-scale. The trade-offs that pathogens make in response to these various environmental pressures add complexity to infectious disease systems and are difficult to study and predict. Models are needed to describe and predict the dynamics of these systems, to develop effective environmental monitoring plans, and to optimize risk-reduction interventions.

Environmental transmission of enteric pathogens depends on a number of factors, including pathogen \textit{persistence} in the environment, i.e., how long the pathogen remains viable outside the host, and pathogen \textit{infectivity}, i.e., the probability of host infection given exposure to the pathogen. We argue that trade-offs between environmental persistence and infectivity, driven by evolutionary or environmental pressures, are likely: a pathogen that evolves to persist in the environment, while maintaining its ability to infect hosts, would enjoy greater transmission and quickly become the dominant strain. Since there is wide variation in persistence times among pathovars, it is likely that longer persistence is accompanied by decreased infectivity.

Microbiological research in infectious disease systems has largely emphasized the identification of genes, gene expression, and metabolic pathways that are associated with virulence, but this work has not translated well into better understanding of risk and disease dynamics at the population level. Although virulence is important to public health, it is infectivity that primarily drives transmission and is therefore integral to risk assessment and control. Yet, the trade-off between persistence and infectivity has not received the same attention from experimentalists as trade-offs involving virulence. Here we examine the dynamical properties associated with the trade-off between persistence and infectivity and the implications for future microbiological work and improved environmental monitoring strategies.

\subsection{Variation in persistence times and infectivity across pathovars and within a pathogen population}

There are clear variations in persistence times and infectivity across closely related pathogen species or pathovars, such as is seen for Escherichia and Salmonella genera. E. coli, in particular, is an extraordinarily diverse group, with only about 6\% of gene families represented in every genome~\citep{Lukjancenko2010}. \textit{E. coli} includes several pathovars that can cause enteric disease, including enterohemorrhagic (EHEC), enteropathogenic (EPEC), enterotoxigenic (ETEC), enteroaggregative (EAEC), and enteroinvasive (EIEC)/\textit{Shigella} types; pathovar designation is largely determined by the presence or absence of virulence genes~\citep{Donnenberg2002,Donnenberg2013}.  Microbial research on pathovars has largely focused on genes, gene expression, and metabolism, and there has been little investigation of how these factors translate into environmental persistence rates and ability to cause infection. Hence, although E. coli pathovars differ in their infectivity and persistence~\citep{Haas2014}, these differences are not well characterized as a whole. There are clues, however, as to how trade-offs at the genetic and metabolic level can propagate to persistence and infectivity. Environmental persistence, for example, may be driven by genes coding for resistance to specific stressors (e.g., resistance to higher temperatures, differences in pH or salinity)~\citep{vanElsas2011}, for pathways to utilize alternate energy sources~\citep{Durso2004}, or the ability to infect and survive in amoebas or other protozoa~\citep{King1988,Lambrecht2015}. Infectivity, on the other hand, may be dependent on whether the infection mechanism acts locally or systemically in the host~\citep{Schmid-Hempel2007,Leggett2012}, as well as the effectiveness of the infection mechanism. Ultimately, evolutionary pressure may direct genetic and metabolic trade-offs across pathovars, resulting in a spectrum of persistence--infectivity strategies.

Variations in persistence times and infectivities can be seen not only across pathovars but also within a population composed of a single pathovar. Variation in persistence, in particular, can be observed as biphasic decay, i.e., long-tailed deviations from the expected monophasic exponential pathogen decay~\citep{Brouwer2017a}. Biphasic decay is well-documented in \textit{E. coli}~\citep{Easton2005,Hellweger2009,Rogers2011,Hofsteenge2013,Zhang2015}, for instance.  While the mechanisms of biphasic decay are not well-understood, hypotheses include population heterogeneity, hardening-off, and the existence of dormant states, such as viable-but-not-cultivable (VBNC)~\citep{Oliver2010,Ramamurthy2014,Li2014} or antibiotic-resistant persister~\citep{Balaban2004,Lewis2007,Maisonneuve2014} states. Whether the extended persistence of pathogens presents a significant public health risk remains an open question, and many risk assessments do not account for it. While it is likely that the persistent subpopulation must sacrifice all or part of its infectivity (at least until it finds more favorable conditions), experimental verification is lacking. We previously developed a mechanistic mathematical framework to describe biphasic decay, both in sampling studies and in quantitative microbial risk assessments~\cite{Brouwer2017a}. Here we extend that work by explicitly examining the dynamic properties of a persistence--infectivity trade-off. Specifically, we i) use an existing transmission model to consider the different outbreak dynamics that can be seen across pathogens making a variety of persistence--infectivity trade-offs and ii) present a new model to consider how within-population phenotypic heterogeneities can further affect outbreak dynamics.

\subsection{The role of disease surveillance in uncovering mechanism}

Direct measurement of the rates of biological processes or other mechanistic parameters through experimental studies is one way to approach these scientific questions. However, sometimes experiments are inconvenient, expensive, or (especially in the case of pathogen challenge studies to determine infectivity) ethically fraught. Indirect methods---using the observation of disease dynamics (i.e., longitudinal disease surveillance) to determine what biological parameters must have been---have played an important part in disease epidemiology. However, as useful and important as disease surveillance is, it does not always contain enough information to untangle more complex underlying mechanisms. The field of identifiability has developed methods to determine which parameters can be uniquely estimated for a model from a given kind of data. This information can then be used to ascertain which experiments or new data collection will have the most power for improving inference. Here, we use identifiability analysis to highlight the ways in which targeted experimental studies could elucidate underlying mechanisms and improve risk assessment and disease control practices.

\section{Models and methods}
\subsection{Models}
\begin{figure}[htbp]
\centering
    \includegraphics[width=1\textwidth]{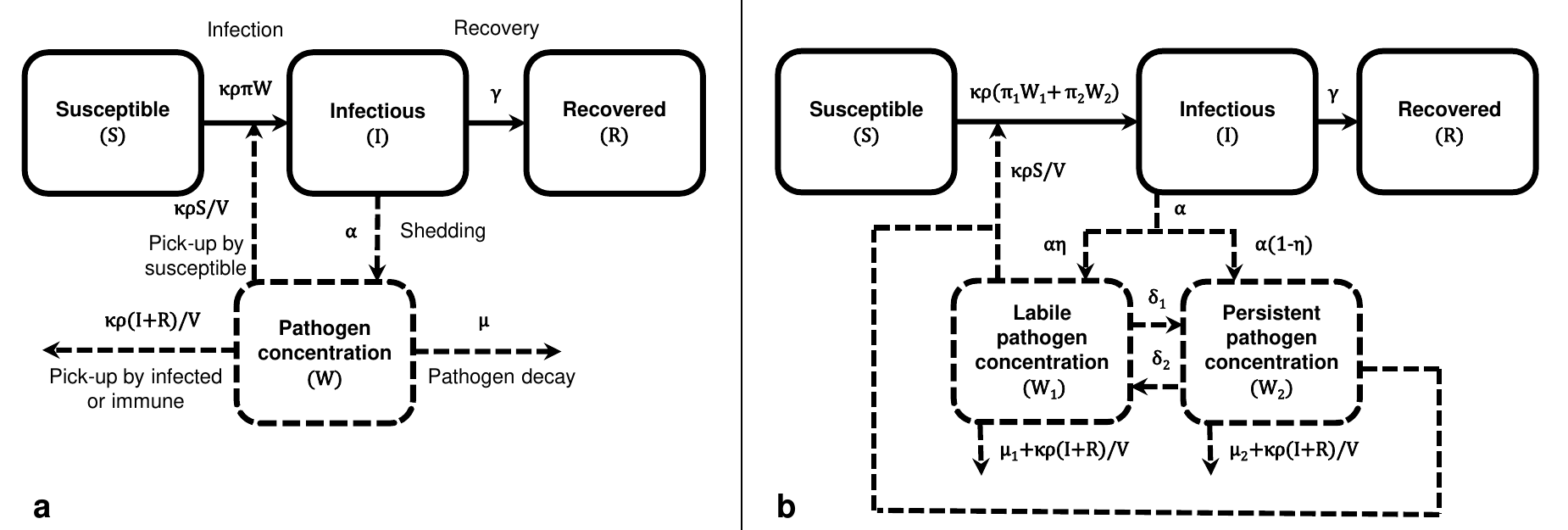}
    \caption{\textbf{Environmentally mediated infectious disease transmission models.} Schematics for models with a) monophasic pathogen decay and b) biphasic pathogen decay, where the pathogen population is comprised of a more infectious, less persistent labile fraction and a less infectious, more persistent fraction.}
    \label{schem}
\end{figure}

We use an environmentally mediated infectious disease transmission model based on a susceptible--infectious--recovered (SIR) framework where all transmission occurs via an environmental compartment and there is no direct person-to-person transmission~\citep{Li2009,Tien2010}.   Model variables and parameters are given in Table~\ref{params}, and a schematic is given in Fig.~\ref{schem}a.
\begin{linenomath}
\begin{align}
\begin{split}
\dot S&= - \kappa \rho  \pi S W,\\
\dot I&=  \kappa  \rho \pi S W-\gamma I,\\
\dot R &= \gamma I,\\
\dot W&= \alpha I - \xi W.
\end{split}
\label{EITS}
\end{align}
\end{linenomath}

In this first model, all pathogens in the population have the same \textit{infectivity} (per-pathogen infection probability $\pi$). Moreover, all pathogens decay with the same monophasic exponential rate ($\xi$), leading to the same average \textit{persistence}, that is, average number of days until removal from the system
\begin{equation}
\tau=\frac{1}{\xi}.
\end{equation}

We make two simplifications from prior environmentally mediated infectious disease models. First, while Li et al.~\citep{Li2009} separately considered the pathogen removal rate from the environment to be the sum of ingestion~$(\kappa\rho N/V)$ and decay~$(\mu)$, we assume that the environmental volume is large enough that the effect of pathogen ingestion on pathogen concentration will be negligible, i.e., we
assume $\kappa\rho N/V\ll \mu$ so that $\xi=\kappa\rho N/V+\mu\approx \mu$.  Li et al.~\citep{Li2009} demonstrated that this assumption leads to density-dependent transmission (as opposed to frequency-dependent transmission). Second, while we have previously considered environmentally mediated infectious disease models with dose--response functions or an exposed individual compartment~\citep{Brouwer_DR}, these aspects are not important for the persistence--infectivity trade-offs we are considering. We do not include them here. 

We extend the above model to account for biphasic decay~\citep{Brouwer2017a} and allow for heterogeneities in population persistence and infectivity. In particular, we assume that the population consists of two subpopulations, one that is more infectious but less persistent (labile pathogens $W_1$) and one that is less infectious but more persistent (persistent pathogens $W_2$). These subpopulations differ in phenotype (gene expression or metabolism) rather than genotype (DNA sequence). Model variables and parameters are given in Table~\ref{params}, and a schematic is given in Fig.~\ref{schem}b. This model allows us to consider the state in which pathogens are shed into the environment (either $W_1$ or $W_2$, determined by $\eta$) and the possibility of phenotype conversion from labile to persistent and from persistent to labile ($\delta_1$ and $\delta_2$, respectively). One or more of these parameters might be negligibly small in practice.
\begin{linenomath}
\begin{align}
\label{EITS_bi}
\begin{split}
\dot S&= - \kappa\rho  S(\pi_1W_1+\pi_2W_2),\\
\dot I&=  \kappa \rho S(\pi_1W_1+\pi_2W_2)-\gamma I,\\
\dot R &= \gamma I,\\
\dot W_1&= \alpha \eta I + \delta_2 W_2 - \left( \xi_1  +\delta_1\right)W_1,\\
\dot W_2&= \alpha (1-\eta) I + \delta_1 W_1 - \left( \xi_2+\delta_2 \right)W_2 .
\end{split}
\end{align}
\end{linenomath}

The two subpopulations can have different \textit{infectivities} ($\pi_1$ and $\pi_2$). The average \textit{persistence} $\tau_i$ for each pathogen type is average amount of time a pathogen stays in a compartment and now includes both removal by decay or pick-up ($\xi$) and phenotypic conversion ($\delta$):
\begin{linenomath}
\begin{equation}
\tau_i=\frac{1}{\xi_i+\delta_i}.
\end{equation}
\end{linenomath}

When the subpopulations have the same infectivity and removal rates ($\pi_1=\pi_2$, $\xi_1=\xi_2$), the model simplifies to the monophasic model above (Eq.~\eqref{EITS}).

\begin{table}
\centering
\caption{\textbf{Variables and parameters of the environmentally mediated infectious disease models.} Models incorporate monophasic pathogen decay~(Eqs.~\eqref{EITS}, Figure~\ref{schem}a) or biphasic decay (Eqs.~\eqref{EITS_bi}, Figure~\ref{schem}b).}
\begin{tabular}{l l}
\hline
{\bf Variables} &\\
$S(t)$ & Number of susceptible people\\
$I(t)$ & Number of infectious people\\
$R(t)$ & Number of recovered people\\
$W(t)$ & Concentration of pathogens in the environment\\
$W_1(t)$ & Concentration of labile pathogens in the environment\\
$W_2(t)$ & Concentration of persistent pathogens in the environment\\
{\bf Parameters} &\\
$\gamma$ & Recovery rate (per day)\\
$\alpha$ & Deposition rate of pathogens per unit volume of environment (per day)\\
$\eta$ & Fraction of deposited pathogens that are labile\\
$\delta_i$ & Rate at which pathogen of phenotype $i$ convert to the other phenotype (per day)\\
$\pi_i$ & Per-pathogen probability of infection for phenotype $i$\\
$\mu_i$ & Pathogen decay rate for phenotype $i$ (per day)\\
$\kappa$ & Rate at which individuals contact the environment (per day)\\
$N$ & Population size\\
$\rho$ & Volume of environment consumed (per contact)\\
$V$ & Total volume of the environment\\
$\xi_i$ & Overall pathogen removal rate for phenotype $i$ (per day), $\mu_i+\kappa\rho N/V$\\
$\tau_i$ & Average persistence of pathogen phenotype $i$ (days)\\
$\phi_i$ & Probability that conversion from phenotype $i$ occurs before decay
\end{tabular}
\label{params}
\end{table}

\clearpage
\subsection{Basic reproduction number}
The \textit{basic reproduction number} $\mathcal{R}_0$, defined as the average number of secondary cases arising from a typical primary case in an entirely susceptible population,  is often used for its epidemic threshold properties, i.e., for initial conditions near the disease-free equilibrium, there will be an epidemic if $\mathcal{R}_0>1$, and the disease will die out if $\mathcal{R}_0<1$. The basic reproduction number is also used to determine needed intervention coverage to eliminate transmission and to estimate the final size of an epidemic. Here, we will use $\mathcal{R}_0$ as a proxy for pathogen fitness, investigating different persistence and infectivity combinations that have the same $\mathcal{R}_0$. We calculate $\mathcal{R}_0$ for our ODE models using the Next Generation Method~\citep{Diekmann1990,vandenDriessche2002}. For the models presented here, the basic reproduction number determines the epidemic attack ratio.

\subsection{Parameter identifiability: estimation and dynamic invariants}
\textit{Identifiability} is the study of what parametric information for a model is available in data. A model parameter is said to be identifiable if its value may be uniquely recovered from the observed data. More formal definitions may be found elsewhere~\citep{Bellman1970, Rothenberg1971,Cobelli1980}. Identifiability analysis is a necessary precursor to parameter estimation as we cannot estimate the value of a parameter if multiple values all give the same output. Parameters that are indistinguishable when measuring one output, however, might be separable for a different output. Identifiability, therefore, is a property of the model together with the specified measurement or data.

Structural identifiability analysis is a theoretical consideration of parameter estimation problems inherent to the structure of the model. For example, one could never separately and uniquely estimate $m_1$ and $m_2$ in the linear model $y=(m_1+m_2)x+b$, no matter how many $(x,y)$-pair data points were measured. Unlike in this example, determination of these sorts of structural limitations is non-trivial for models of even modest complexity. Structural identifiability analysis is contrasted with practical identifiability analysis, which considers problems arising in parameter estimation of real-world data~\citep{Raue2009}. In the rest of this manuscript, we will use ``identifiable'' to mean ``structurally identifiable,'' unless otherwise specified.

If not all parameters can be identified from data, one can find algebraic combinations of the parameters that are identifiable from the data~\citep{Cobelli1980}. These \textit{identifiable parameter combinations} are central to identifiability analysis and the model dynamics: any set of individual parameter values with the same value of their algebraic combination will produce the same dynamics. This means that the identifiable parameter combinations are invariants for the dynamics.

To compute the identifiable parameter combinations, we use a differential algebra approach to identifiability~\citep{Saccomani2001, Audoly2001, Meshkat2009}, finding an input--output equation, which is a monic, polynomial equation that can be written in terms of only the observed state variable (i.e., the data variable), its derivatives, and the model parameters. The input--output equation has equivalent observed dynamics to that of the original ODE system, and the coefficients of the input--output equation are the identifiable parameter combinations. Mathematical details and proofs are left to the supplement.

\subsection{Computation}
Integration of ODE models was done in R (v3.4.1) with the \texttt{deSolve} package, and parameter estimation was done using a David--Fletcher--Powell algorithm in the \texttt{Bhat} package. The differential algebra computation for the identifiability analysis was done in Mathematica (v11.1).

\section{Results}

This first section presents the basic reproduction number and the identifiable parameter combinations for both the monophasic and biphasic pathogen decay models. We next examine the how the persistence--infectivity trade-off affects the transmission dynamics, and finally we elucidate what data are need to fully identify these environmentally mediated transmission models.

\subsection{The basic reproduction number and identifiability}
\subsubsection{Monophasic decay disease model} 
The basic reproduction number for the model with monophasic decay is given by~\citep{Li2009,Tien2010}
\begin{linenomath}
\begin{equation}
\mathcal{R}_0= %\frac{\alpha \pi\kappa\rho N}{\gamma\xi}=
\frac{\alpha \pi\kappa\rho\tau N}{\gamma}.
\label{R0_mono}
\end{equation}
\end{linenomath}

The identifiable combinations of this model have been published elsewhere~\citep{Eisenberg2013}. In brief, if case data (corresponding to state $I$) is observed, we can uniquely determine the recovery rate $\gamma$, the average pathogen persistence $\tau$, and the product $\alpha\pi\kappa\rho$. The basic reproduction number $\mathcal{R}_0$, therefore, is structurally identifiable if the population size $N$ is known.

The parameter combination $\alpha\pi\kappa\rho$ can be understood in the following way. The product $\kappa\rho$ is the volume of the environment ingested per day, and $\pi$ is the per pathogen probability of infection. Then, $\pi\kappa\rho$ is the rate of infection-transmitting contact with the environment. The shedding rate ($\alpha$) and infectious contact rate ($\pi\kappa\rho$) are in an identifiable parameter combination when we only observe infections in the population ($I$). In this case, we do not measure the concentration of pathogens in the environment ($W$), and we do not know whether the force of infection ($\pi\kappa\rho W$) is a result of fewer pathogens that have a higher rate of infection (low shedding $\alpha$, high rate of infectious contact $\pi\kappa\rho$) or more pathogens with a lower rate of infection (high shedding $\alpha$, low rate of infectious contact $\pi\kappa\rho$). If the concentration of pathogens in the environment is observed, however, the relative sizes of the shedding rate ($\alpha$) and infectious contact rate ($\pi\kappa\rho$) are distinguishable. That is, if environmental monitoring data ($W$) is also available in addition to case data ($I$), then $\alpha$ and $\pi\kappa\rho$, not just their product, are separately identifiable.

\subsubsection{Biphasic decay disease model}
To interpret the basic reproduction number $\mathcal{R}_0$ and the identifiable parameter combinations of the biphasic pathogen decay model~(Eqs.~\eqref{EITS_bi}), we introduce two new variables. 
Pathogens leave their environmental compartment either by decay $(\xi_i)$ or by phenotype conversion $(\delta_i)$, and we denote the probability that conversion occurs before decay by 
\begin{linenomath}
\begin{equation}
\phi_{i}:=\frac{\delta_i}{\xi_i+\delta_i}.
\end{equation}
\end{linenomath}

Then, we may write and interpret the basic reproduction number $\mathcal{R}_0$ as follows.

\begin{proposition}
The basic reproduction number for the environmentally mediated infectious disease transmission model with biphasic pathogen decay~(Eqs.~\eqref{EITS_bi}) is 
\begin{linenomath}
\begin{equation}
\mathcal{R}_0= \frac{\alpha\kappa\rho N}{\gamma}\left(\pi_1\tau_1\left(\frac{\eta+(1-\eta)\phi_2}{1-\phi_1\phi_2}\right) + \pi_2\tau_2\left(\frac{(1-\eta)+\eta \phi_1}{1-\phi_1\phi_2}\right)\right).
\label{R0_bi}
\end{equation}
\end{linenomath}
\end{proposition}

The proof is left to the supplement. This system $\mathcal{R}_0$ can be seen as the sum of two submodel basic reproduction numbers that give the contributions of the labile ($\mathcal{R}_{0,1}$) and persistent ($\mathcal{R}_{0,2}$) subpopulations to the overall basic reproduction number.
\begin{linenomath}
\begin{align}
\begin{split}
\mathcal{R}_0&= \left( \frac{\alpha\pi_1\kappa\rho\tau_1 N }{\gamma}\right) \left(\frac{\eta+(1-\eta)\phi_2}{1-\phi_1\phi_2}\right)+\left( \frac{\alpha\pi_2\kappa\rho\tau_2 N}{\gamma}\right)\left(\frac{(1-\eta)+\eta \phi_1}{1-\phi_1\phi_2}\right),
\end{split}\\
&=:\mathcal{R}_{0,1}+\mathcal{R}_{0,2}.
\end{align}
\end{linenomath}

These two submodels are each similar in form to the monophasic basic reproduction number (Eq.~\eqref{R0_mono}), with a coefficient that accounts for the interconnectedness of the two compartments.  Of the $\alpha$ pathogens shed between the two compartments, $\alpha\eta$ go directly to the labile compartment, but $\alpha(1-\eta)\phi_2$ will also come to the labile compartment via the persistent compartment. Because pathogens can move back and forth between compartments, we need to know  the expected number of visits a pathogen makes to the labile compartment~\citep{vandenDriessche2002}.  After the initial visit, each return visit happens with probability $\phi_1\phi_2$. Thus, the expected amount of time spent in the labile compartment is
\begin{linenomath}
\begin{equation}
\tau_1(1+\phi_1\phi_2+(\phi_1\phi_2)^2+\cdots+(\phi_1\phi_2)^n+\cdots )= \frac{\tau_1}{1-\phi_1\phi_2}.
\end{equation}
\end{linenomath}

Because these submodel reproduction numbers allow us to understand the relative contribution of each pathogen phenotype to the overall epidemic potential of the system, we would like to be able to determine their values from data.  In particular, we want to understand the risk potential in the less infectious, persistent fraction of pathogens. To this end, we compute the identifiable parameter combinations for the biphasic decay model to inform what data are required to provide useful information from the model.

\begin{proposition}
\textbf{(a).} The structurally identifiable parameter combinations for the environmentally mediated infectious disease transmission model with biphasic pathogen decay~(Eqs.~\eqref{EITS_bi}) if case data ($I$) are observed are
\[\gamma,\]
\[\alpha(\eta\pi_1+(1-\eta)\pi_2)\kappa\rho,\]
\[\xi_1+\delta_1+\xi_2+\delta_2=\frac{\tau_1+\tau_2}{\tau_1\tau_2},\]
\[(\xi_1+\delta_1)(\xi_2+\delta_2)-\delta_1\delta_2=\frac{1-\phi_1\phi_2}{\tau_1\tau_2},\]
\[\mathcal{R}_0/N.\]
\label{bi_ident_comb}
\end{proposition}

In this case, the disease recovery rate $\gamma$ is identifiable as it was in the monophasic decay model. The combination $\alpha(\eta\pi_1+(1-\eta)\pi_2)\kappa\rho$ in the biphasic model has an analogous interpretation to that of the combination $\alpha\pi\kappa\rho$ in the monophasic model.
The two identifiable parameter combinations $\xi_1+\delta_1+\xi_2+\delta_2$ and $(\xi_1+\delta_1)(\xi_2+\delta_2)-\delta_1\delta_2$ come directly from the underlying biphasic pathogen decay model that we previously described in Brouwer et al.~\citep{Brouwer2017a}; they are the sum and product of the apparent labile and persistent decay rates. Because $\mathcal{R}_0/N$ is identifiable from case data, the basic reproduction number will be identifiable if the population size is known.

\setcounter{ordprop}{1}
\begin{proposition}
\textbf{(b).} 
If the total pathogen concentration is also observed ($W=W_1+W_2$), then 
\[\alpha,\]
\[\frac{ \tau_1(\eta+(1-\eta)\phi_2)+\tau_2((1-\eta)+\eta \phi_1)}{\tau_1\tau_2},\]

are also structurally identifiable. If the persistant pathogens are not culturable by usual methods and only labile pathogen concentration ($W_1$) is observed, then the additional identifiable parameter combinations are instead
\[\alpha\eta,\]
\[\frac{1}{\eta}\frac{\tau_1(\eta+(1-\eta)\phi_2)}{\tau_1\tau_2}.\]
One caveat to this last result is that we cannot distinguish $W$ from $W_1$ using case and environmental monitoring data alone. If the labile ($W_1$) and persistent ($W_2$) pathogens can be separately quantified, then $\gamma$, $\alpha$, $\eta$, $\delta_1$, $\delta_2$, $\xi_1$, $\xi_2$, $\kappa\rho\pi_1$, and $\kappa\rho\pi_2$ are structurally identifiable.
\end{proposition}

The proof is left to the supplement. Proposition~\ref{bi_ident_comb} suggests that by separately quantifying the labile and persistent pathogens, we can almost fully determine the parameters, though there may be practical barriers for real-world data. The other option is to combine case or prevalence data ($I$). The other option is to combine case or prevalence data ($I$) and total pathogen concentration ($W=W_1+W_2$) with \textit{a priori} parameter values to extract more information out of the identifiable parameter combinations.

\subsection{Persistence--infectivity trade-offs affect outbreak dynamics}
\begin{figure}[htbp]
\centering
    \includegraphics[width=1\textwidth]{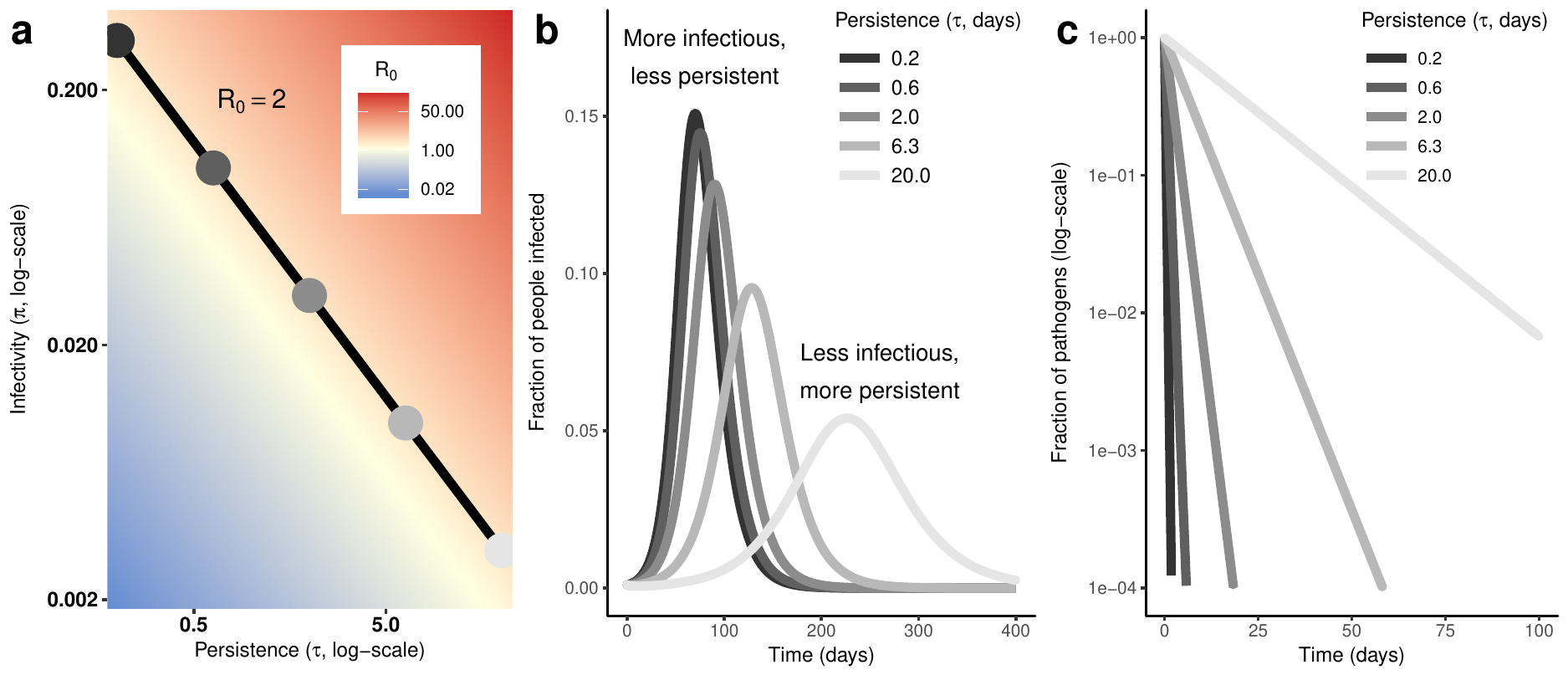}
    \caption{\textbf{Infectivity--persistence trade-offs in a monophasic pathogen decay model.} (a).~Heatmap of the basic reproduction number $\mathcal{R}_0$ of the monophasic decay disease model (Eqs.~\eqref{EITS}) as a function of persistence $(\tau)$ and infectivity $(\pi$). The line is the contour along which $\mathcal{R}_0=2$. Here, $N=$ 1000, $\gamma=$  0.1, $\kappa=$  8, $\rho=$ 0.15, $\alpha=$ 0.001. The colored dots correspond to the colored lines in (b) and (c). (b)~Fraction of the population infected for the values of pathogen persistence $(\tau)$ and infectivity $(\pi)$ given by the dots in (a). Although all points have $\mathcal{R}_0=2$, the epidemic dynamics vary significantly over the individual parameter values. (c)~Pathogen decay curves in the absence of a system input illustrate the variation in the corresponding persistences $(\tau)$.}
    \label{PIT_mono}
\end{figure}

The pathogen infectivity $\pi$ and the pathogen persistence $\tau$ are not part of the same identifiable parameter combinations in the monophasic decay model. This means that differences in the outbreak dynamics can be observed when we compare a highly infectious pathogen with low persistence to a less infectious pathogen with high persistence. At the same time, infectivity $\pi$ and persistence $\tau$ appear in a product in the analytic equation for $\mathcal{R}_0$ (eq.~\eqref{R0_mono}).  As long as the product of $\pi$ and $\tau$ is constant, the basic reproduction number, and, therefore, the attack ratio, will be the same. Altogether, the persistence--infectivity trade-off can produce a variety of dynamics all associated with the same $\mathcal{R}_0$, and we find that slower outbreaks with smaller peak sizes are associated with less infectious, more persistent pathogens (Figure~\ref{PIT_mono}).

\begin{figure}[htbp]
\centering
    \includegraphics[width=1\textwidth]{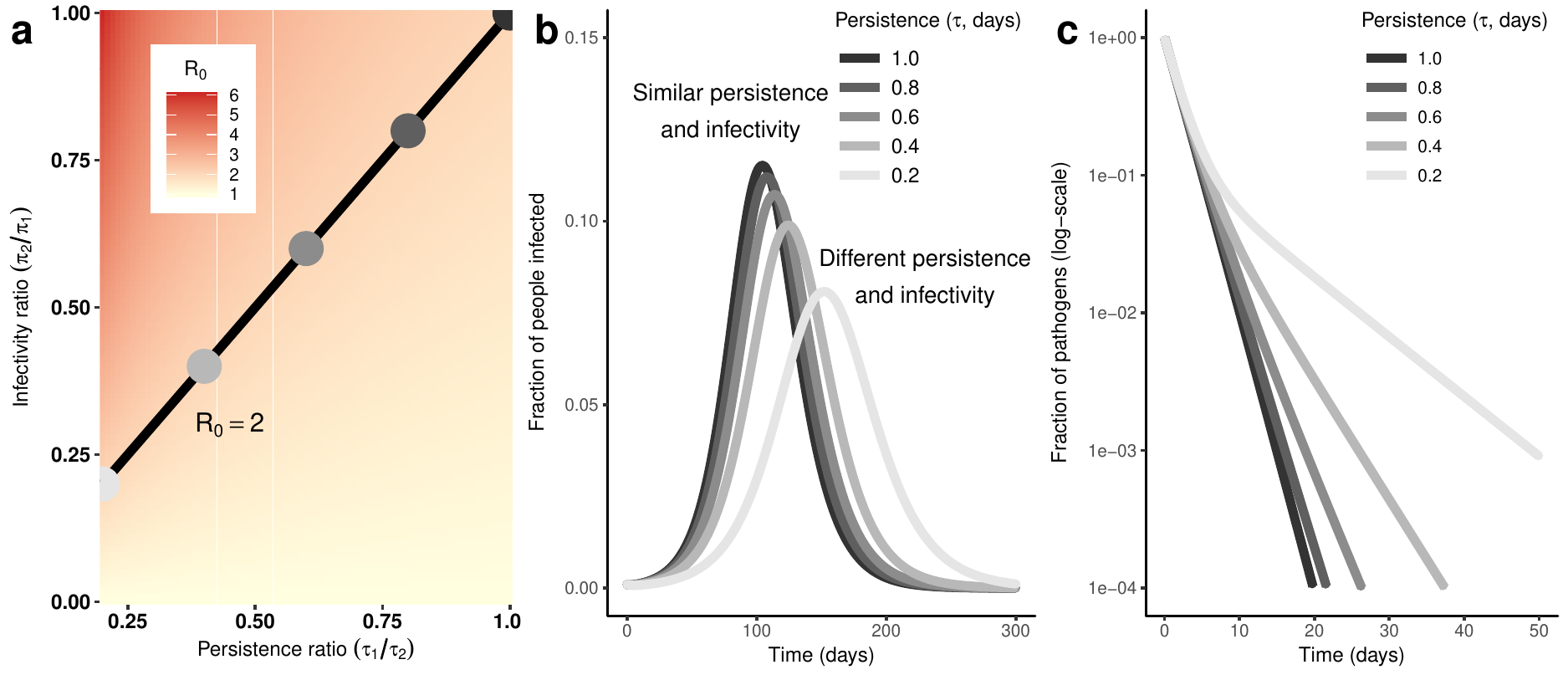}
    \caption{\textbf{Infectivity--persistence trade-offs in a biphasic pathogen decay model.} (a)~Heatmap of the basic reproduction number $\mathcal{R}_0$ of the biphasic decay disease model (Eqs.~\eqref{EITS_bi}) as a function of the ratio of the persistences $(\tau_1/\tau_2)$ and infectivities $(\pi_2/\pi_1)$. The line is the contour along which $\mathcal{R}_0=2$.  The colored dots correspond to the colored lines in (b) and (c). (b)~ Fraction of the population infected over time for the values of pathogen persistence and infectivity ratios given by the dots in (a). Although all points have $\mathcal{R}_0=2$, the epidemic dynamics vary significantly over the parameter ratios. Here, $N=$ 1000, $\gamma=$  0.1, $\kappa=$  8, $\rho=$ 0.15, $\alpha=$ 0.001, $\eta=$ 0.99, $\phi_1=$ 0.1, $\phi_2=$ 0.01, $\pi_1=0.0195$, $\tau_1=2$. (c)~Pathogen decay curves in the absence of a pathogen input illustrate the degree of biphasic behavrior corresponding to the persistence ratios $(\tau_1/\tau_2)$.}
    \label{PIT_bi}
\end{figure}

In the biphasic decay disease model, we observe the same phenomena. Heuristically, the trade-offs are more easily observed if we express the degree of deviation from monophasic behavior using the ratios of infectivities $\pi_2/\pi_1$ and persistence times $\tau_1/\tau_2$, where more deviation from 1 indicates a greater deviation from monophasic behavior. Because we consider only the case where the persistent subpopulation is no less persistent and no more infectious than the labile subpopulation, we only consider $0<\pi_2/\pi_1<1$ and  $0<\tau_1/\tau_2<1$. Rewriting the basic reproduction number in terms of these ratios,
\begin{linenomath}
\begin{equation}
\mathcal{R}_0= \frac{\alpha\kappa\rho \pi_1\tau_1 N}{\gamma}\left(\left(\frac{\eta+(1-\eta)\phi_2}{1-\phi_1\phi_2}\right) + \frac{\pi_2/\pi_1}{\tau_1/\tau_2}\left(\frac{(1-\eta)+\eta \phi_1}{1-\phi_1\phi_2}\right)\right),
\end{equation}
\end{linenomath}

we see that when we fix the persistence and infectivity of the labile subpopulation ($\tau_1$ and $\pi_1$),  the two ratios  $\pi_2/\pi_1$ and $\tau_1/\tau_2$ must be  proportional to maintain $\mathcal{R}_0$
(Figure~\ref{PIT_bi}a).  As with the monophasic model, the outbreak peaks later and smaller as the persistent pathogens become relatively more persistent but less infectious,  (Figure~\ref{PIT_bi}b). At the same time, the biphasic deviation become more pronounced (Figure~\ref{PIT_bi}c). 

Pathogens can fall in different places along the persistence--infectivity spectrum while still maintaining the same basic reproduction number, a proxy for pathogen fitness and measure of the attack ratio. Moreover, two phenotypes within a single population might likewise have different persistence--infectivity strategies, and thereby exhibit biphasic decay. In both the monophasic and biphasic decay disease models, these persistence--infectivity trade-offs have implications for the timing and peak size of the associated epidemics, which may in turn direct control strategies.

\subsection{The need for both disease surveillance and parameter data to elucidate mechanism}

In this section, we demonstrate that disease incidence data can be used in conjunction with experimental studies to improve inference of model parameter values. This form of model parameter identification can guide control efforts and significantly improve risk assessment practices.

\begin{figure}[p]
\centering
    \includegraphics[width=0.85\textwidth]{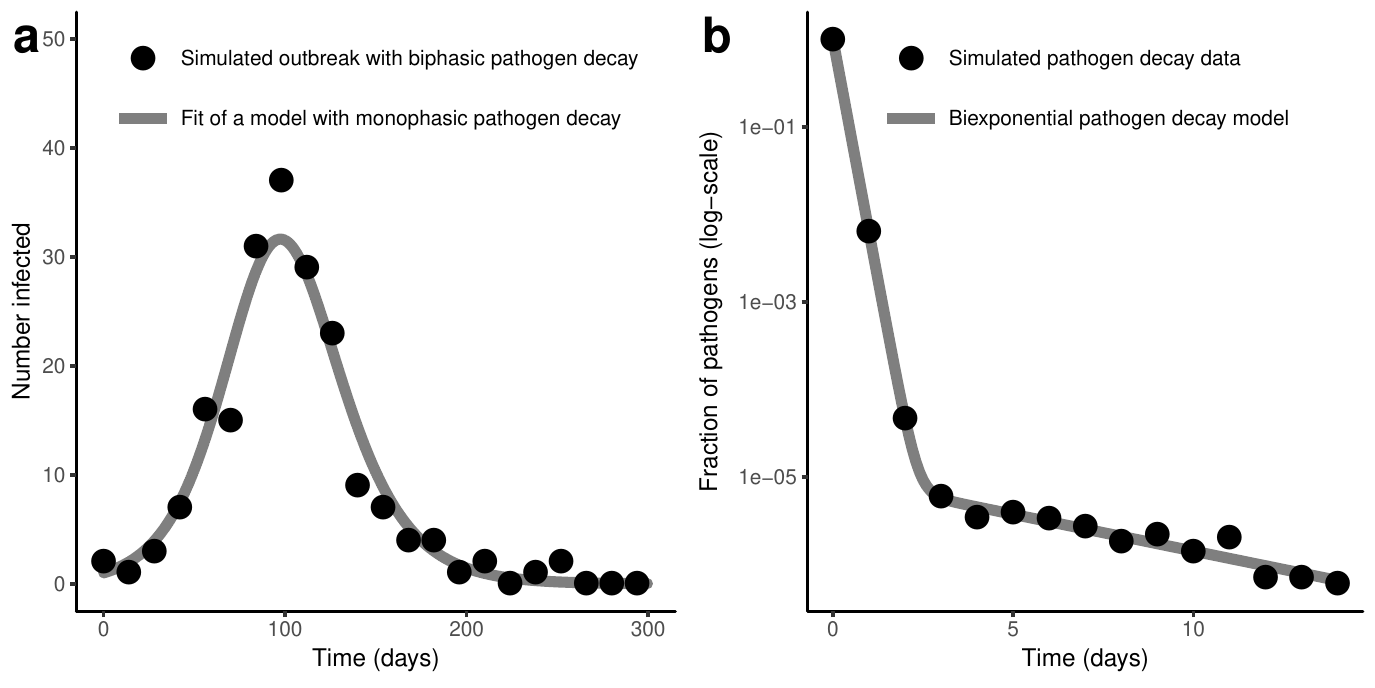}
    \caption{\textbf{Pathogen decay dynamics cannot be inferred from case data alone.} (a)~An outbreak of {\it Shigella} simulated with the biphasic decay disease model with parameters $N=$ 1000, $\gamma=$  1/6, $\kappa=$  8, $\rho=$ 0.15, $V=$ 4E8, $\eta=$ 1-(1E-5), $\pi_1=$ 1.1E-2, $\pi_2=$ 1.1E-4, $\xi_1=$ 5, $\xi_2=$ 0.2, $\delta_1=$ 0.05, $\delta_2=$ 0.002, and $\mathcal{R}_0=$ 1.3. Biweekly case data were simulated from a binomial distribution. The monophasic decay disease model was fit to this simulated data using a binomial likelihood: parameter estimates are $\alpha\kappa\rho\pi=$ 3.4E-3 (95\% CI: 2.8E-4, 4.1E-2), $\gamma=$ 1.6E-1 (95\% CI: 1.5E-1, 1.8E-1), $\tau=$ 6.3E-2 (95\% CI: 5.2E-3, 7.6E-1), $I(0)=$1.0 (95\% CI: 0.6,1.8). The aymptotic confidence intervals for $\alpha\kappa\rho\pi$ and $\tau$ are very wide. (b)~Simulated pathogen decay data reveals a biphasic decay pattern and allows the estimation of the apparent fast and slow decay rates. The sum and product of these rates are represented by $\xi_1+\delta_1+\xi_2+\delta_2=$ 5.3 and $ (\xi_1+\delta_1)(\xi_2+\delta_2)-\delta_1\delta_2$=  0.98. Using these estimates in fitting the biphasic decay disease model to the data allows us to resolve the practical identifiability issue and make more accurate inferences: $\alpha\kappa\rho(\eta\pi_1+(1-\eta)\pi_2)$ (true: 1.09E-3, estimated: 6.22E-4) and $\mathcal{R}_0/N$ (true: 1.30E-3, estimated: 1.44E-3).}
    \label{Shigella}
\end{figure}

To this end, we simulate an outbreak of {\it Shigella} using the full biphasic decay disease model (Eqs~\eqref{EITS_bi}) for a small village on a lake and consider the perspective of a researcher trying to use disease surveillance to elucidate the underlying dynamics. We fit the monophasic decay disease model (Eqs.~\eqref{EITS}) to the biweekly surveillance data (Figure~\ref{Shigella}a). The monophasic environmentally mediated infectious disease transmission model captures the dynamics of the biphasic data well; i.e. we cannot detect any model misspecification from the fit alone. Even though we capture the dynamics, our parameter estimates are highly uncertain. Estimates of both $\alpha\kappa\rho\pi$ and $\tau$ vary by two orders of magnitude~(Figure~\ref{Shigella} legend). In particular, the persistence $\tau$ can be arbitrarily small and still fit the data well. 

To improve our parameter inference, we could experimentally determine the environmental persistence of the pathogen in this context. We conduct an experimental pathogen decay study, taking a sample of recently shed pathogen and observing its decay in a controlled environment (simulated Figure~\ref{Shigella}b). The pathogen decay study indicates that the pathogen decay is actually biphasic, which we did not detect from disease surveillance data alone. Fitting a biexponential model to this data allows us to estimate the sum and product of the apparent fast and slow decay rates ($\xi_1+\delta_1+\xi_2+\delta_2$ and $ (\xi_1+\delta_1)(\xi_2+\delta_2)-\delta_1\delta_2$, respectively~\citep{Brouwer2017a}.

We can use this experimentally derived data in the parameter estimation for the biphasic decay disease model to circumvent the inference problems and estimate $\gamma$, $\alpha(\eta\pi_1+(1-\eta)\pi_2)\kappa\rho$, and $\mathcal{R}_0/N$ with more accuracy and precision. The number of parameters (eleven), however, is greater than the number of degrees of freedom in the information (five), meaning that we can say very little about the values of the individual parameters, other than putting some general bounds on $\tau_1$, $\tau_2$, and $\alpha\kappa\rho$. This example highlights how independent experimental determination of parameter values can begin to improve parameter inference when only disease surveillance data is available but that additional parameter information may be needed to further distinguish individual model parameters.

More broadly, a better understanding of the underlying mechanism can be achieved by two experimental strategies. First, experimental studies could be designed to independently identify parameter values. Here, the important parameters to identify are the infectivity of pathogens in the persistent state $(\pi_2)$, the rates of entering dormancy ($\delta_1$) and of resuscitation ($\delta_2$), and the fraction of pathogens already dormant when initially shed into the environment ($1-\eta$)); identifying these parameters is essential to understanding the relative risks associated with the labile and persistent pathogen subpopulations. In particular, determining that one or more of these parameters is negligibly small could significantly simplify the modeling framework.  The second experimental strategy would be to quantify the dynamics of the labile and persistent subpopulations separately, which theoretically resolves nearly all parameters (Proposition~\ref{bi_ident_comb}). Pursuing both experimental strategies simultaneously will allow for corroboration and  maximize our confidence in the conclusions of individual experimental studies because theoretical identifiability does not guarantee that real-world data will contain sufficient information to distinguish the mechanistic parameters in practice

\section{Discussion}
Microbial pathogens make trade-offs to maximize survival. For pathogens that require a human host to reproduce, trade-offs will likely maximize transmission potential. These trade-offs can unfold over a long time-scale (evolutionary) or shorter time-scales (metabolic, gene expression, horizontal gene transfer~\citep{Gluck-Thaler2015}, etc.). With regard to a persistence--infectivity trade-off, our analysis indicates that highly infectious pathogens with low persistence have faster epidemic dynamics than persistent but less-infectious pathogens, even when the total risk is the same. Identification of one piece of this trade-off---variation in persistence times within pathogen subpopulations---will not be possible with disease surveillance alone. Additionally, environmental surveillance of the total number of pathogens, with no distinction between subpopulations with different persistence phenotypes, will also not be sufficient to uniquely estimate the underlying biological parameters. Selected experimental studies to ascertain key parameter values, on the other hand, can maximize the information available in disease and environmental surveillance and lead to fully specified risk models.

\subsection{Risk assessment, biphasic decay, and dormant states}

Currently, risk assessments often do not consider heterogeneity of pathogen populations in water. In the case of waterborne risk of \textit{Helicobacter pylori}, for example, published risk assessments use an infectivity based on a less-persistent, culturable state~\citep{Ryan2014}, despite the fact that \textit{H. pylori} transforms to a more-persistent but less-infectious VBNC state within days of entering water~\citep{Adams2003}.  Characterizing the persistence--infectivity trade-off within pathogen populations in the environment will have important implications for risk assessment, particularly when biphasic decay is possible. Biphasic decay indicates the presence of labile and persistent phenotypes, each with a different associated disease potential (characterized by $\mathcal{R}_{0,1}$ and $\mathcal{R}_{0,2}$, respectively). Understanding how much the more persistent subpopulation contributes to the overall epidemic potential would improve the accuracy of risk assessments. Most risk assessments assume that pathogens decay exponentially (i.e., monophasically), discounting the possibility of a persistent subpopulation. Even when there is experimental evidence to characterize the persistences of the subpopulations, there has so far been little indication of whether to treat the persistent population as comparably infectious, not infectious, or somewhere in between. Experimental studies that provide information on the relative infectivity of the persistent subpopulation, as well as other biological parameters, will be useful for assessing the role of persistent subpopulations in public health risk assessment.

Questions of trade-offs between environmental persistence and infectivity are particularly relevant to the study of dormant microbial states---such as VBNC or antibiotic-resistant persister states---that result from environmental stresses. The VBNC state has been observed in many bacterial species and is characterized by a lack of culturability with classical techniques. Over fifty human pathogens have been reported to exhibit a VBNC state, including \textit{E. coli}, \textit{Salmonella}, and \textit{Vibrio cholerae}~\citep{Oliver2010,Li2014}.  It is thought that the VBNC state is an adaptive strategy for survival in unfavorable environments, is induced by environmental stresses including disinfection, and can be reversed through resuscitation~\citep{Oliver2010,Ramamurthy2014,Li2014}. However, the exact role and mechanisms of the VBNC state appear seem to vary between species, so the VBNC state may be better considered a collection of related but species-specific states that are all characterized by a loss of culturability. For most species, the VBNC state has a slower die-off rate, and there is evidence of reduced infectivity in some species (e.g. \textit{H. pylori}~\citep{Boehnke2017}). Additionally, there is evidence that cells can regain full virulence upon resuscitation~\citep{Oliver2010,Ramamurthy2014,Li2014}, though \textit{in vivo} resuscitation of ingested VBNC pathogens may be comparatively rare~\citep{Winfield2003}.  

Persister states are a similar dormant state observed in the prescence of antibiotics. Persister cells appear to be the same genotype as the rest of the population but have a different phenotype with respect to antibiotic resistance. This resistance comes at the price of non-proliferation~\citep{Balaban2004,Lewis2007}. The presence of these dormant persister cells in biofilms has been implicated in multidrug-tolerant infections~\citep{Lewis2007}. Although some work has been done to understand the underlying molecular mechanisms~\citep{Maisonneuve2014}, questions remain~\citep{Balaban2013}. An emerging hypothesis, called the ``dormancy continuum," has suggested that the VBNC state and persister state are closely related phenomena~\citep{Ayrapetyan2015,Ayrapetyan2015a}.  

Although understanding of VBNC and persister states is increasing, there is currently limited evidence for a link between them and specific mechanistic trade-offs between persistence and infectivity. This may be because experimentalists have not tried to address this specific question, or, it may be because metabolic trade-offs occur at the level of gene expression and only poorly translate into the high-level concepts of persistence and infectivity. Indeed, evidence suggests that increased resistance to environmental decay is stressor-specific, meaning that differences may not be detected if experimental designs are not exploiting the true underlying mechanism.

\subsection{Directions for experimental microbiologists}
Although very important from a public health standpoint, disease surveillance contains only limited information about underlying environmentally mediated pathogen dynamics.  Experimentalists have the opportunity, therefore, to expand not only our knowledge of the biology of human pathogens in the environment, but also, by collaborating with mathematical disease modelers, the implications for population-level disease. Experimentalists can directly ascertain values of biological parameters, including the relative infectivity of persistent pathogens and the rates of phenotypic changes; calls for assessing rates of entering dormancy and resuscitation have already been made by mathematical modelers assessing persister cells~\citep{Hofsteenge2013}. Alternatively, by developing methods that separately quantify the dynamics of labile and persistent pathogens, experimentalists can help provide the time-series data modelers need to indirectly infer the biological parameters and rates. These direct and indirect pathways are complementary, and both should be pursued.

More generally, microbiologists could contribute to public health by focusing on the downstream implications of gene expression and metabolic processes. For example, how does the presence of a certain gene in a pathogen population translate into its ability to infect those exposed to it? Or, under which real-world environmental conditions should we expect extended persistence of still-infectious pathogens? Shifting the experimental mindset to answering such questions will significantly benefit risk assessment and public health researchers.

\subsection{Methodological limitations and advances}

In this analysis, we used the basic reproduction number $\mathcal{R}_0$ as a proxy for pathogen fitness. Although $\mathcal{R}_0$ is likely an acceptable first approximation for pathogen fitness, real-world trade-offs may not precisely maintain the basic reproduction number because of the complex nature of genetic or metabolic trade-offs and because bacteria are not inherently optimizing the abstract concept of human disease transmission (there may be local, contextual effects or the influence of other hosts). Also, by using a compartmental model, we make implicit assumptions about deterministic and well-mixed dynamics. True dynamics are likely to be spatial (perhaps related to biofilm formation) and likely stochastic; indeed, persister cells, for example, are thought to arise from stochastic fluctuations in gene expression and to comprise only a small fraction of the population, on the order of 10\textsuperscript{-4} to 10\textsuperscript{-6}. Nevertheless, our work provides a first mathematical modeling framework for considering possible population-level public health implications of microbial dormancy.

Methodologically, this study also contributes to the field of identifiability. While identifiability is traditionally concerned with parameter estimation from data, here we leverage identifiability to conduct dynamics analysis, emphasizing that identifiable parameter combinations are dynamic invariants; any set of parameters that produces the same values for the set of identifiable parameter combinations will produce the same outbreak trajectory (or other model output). This approach allows us to reduce the model’s parameter space to the number of identifiable parameter combinations, which allows us to more efficiently understand our models and ask targeted questions.

\subsection{Conclusion}

To improve microbial risk assessments of environmentally mediated pathogens and to provide a more precise means of developing environmentally-based control strategies, we will require a better understanding of the dynamic mechanisms that drive the variation in pathogen persistence times, as well as the associated trade-offs with infectivity. The development and analysis of the dynamic models presented here create a framework to translate data obtained from microbial experimental research into predictions of population-level public health outcomes. In particular, targeted experimental studies designed to elucidate the dynamics of persistent pathogen subpopulations are needed.

\section*{Acknowledgments}
This work was funded under the Models of Infectious Disease Agent Study (MIDAS) program within the National Institute of General Medical Sciences of the National Institutes of Health (grant number U01GM110712) and under the National Science Foundation Water Sustainability and Climate Program (grant 1360330). AFB, MCE, JNSE received funding from one or both of these sources. The funders had no role in study design, data collection and analysis, decision to publish, or preparation of the manuscript.

\clearpage
\section*{References}
\scriptsize
\bibliography{Persistence-infectivity}
\bibliographystyle{model4-names}\biboptions{authoryear}

\end{document}